\documentclass[rnote,traditabstract]{aa} 
\usepackage{graphicx}
\usepackage{txfonts}
\usepackage{natbib}
\bibpunct{(}{)}{;}{a}{}{,} 

\begin{document}
   \title{On the Progenitor System of Nova V2491~Cygni}

   \author{M.~J. Darnley\inst{1}\fnmsep\thanks{\email{M.Darnley@astro.ljmu.ac.uk}}
          \and
           V.~A.~R.~M. Ribeiro\inst{1}
          \and
           M.~F. Bode\inst{1}
          \and
          U. Munari\inst{2,3}}

   \institute{Astrophysics Research Institute, Liverpool John Moores University, Egerton Wharf, Birkenhead, CH41~1LD, UK
         \and
             INAF Astronomical Observatory of Padova, via dell'Osservatorio, 36012 Asiago (VI), Italy
         \and
             ANS Collaboration c/o Astronomical Observatory, 36012 Asiago (VI), Italy
             }

   \date{Received November 2, 2010; accepted April 12, 2011}

 
  \abstract{
  Nova V2491~Cyg is one of just two detected pre-outburst in X-rays.  The light curve of this nova exhibited a rare ``re-brightening'' which has been attributed by some as the system being a polar, whilst others claim that a magnetic WD is unlikely.  By virtue of the nature of X-ray and spectroscopic observations the system has been proposed as a recurrent nova, however the adoption of a 0.1 day orbital period is generally seen as incompatible with such a system.  In this research note we address the nature of the progenitor system and the source of the 0.1 day periodicity.  Through the combination of Liverpool Telescope observations with published data and archival 2MASS data we show that V2491~Cyg, at a distance of $10.5 - 14$~kpc, is likely to be a recurrent nova of the U~Sco-class; containing a sub-giant secondary and an accretion disk, rather than accretion directly onto the poles.  We show that there is little evidence, at quiescence, supporting a $\sim0.1$ day periodicity, the variation seen at this stage is likely caused by flickering of a re-established accretion disk.  We propose that the periodicity seen shortly after outburst is more likely related to the outburst rather than the - then obscured - binary system.  Finally we address the distance to the system, and show that a significantly lower distance ($\sim2$~kpc) would result in a severely under-luminous outburst, and as such favour the larger distance and the recurrent nova scenario. 
}

   \maketitle
%

\section{Introduction}

The typical classical nova (CN) system consists of a white dwarf (WD) primary and a main-sequence secondary in a close binary orbit; the secondary star fills its Roche lobe.  Material lost by the secondary accumulates onto the surface of the WD via an accretion disk around the WD which lies in the orbital plane.  In a small number of cases - the {\it polars} - accretion is channelled to the magnetic poles of the WD and no appreciable disk is present.  Nuclear burning can initiate within this accreted, degenerate, material, leading to a thermonuclear runaway which will eject some or all of this accreted layer.  A nova eruption can reach $M_{V}\simeq-10$ \citep{2009ApJ...690.1148S} and may eject $10^{-5}-10^{-4}$~M$_{\sun}$ of matter \citep[e.g.][and references therein]{2010AN....331..160B}.

The closely related recurrent nova (RN) systems exhibit recurrence timescales of $10-100$ years.  This is attributed to a combination of a high mass WD and high accretion rate.  The elevated accretion rate is caused by the presence of an evolved secondary; either a sub-giant (U~Sco-class) or a red giant (RS~Oph-class).  However, the T~Pyx-class of recurrents contain lower mass WDs and main sequence secondaries, but other than their short recurrence timescales, they are more akin to CNe \citep{2008ASPC..401...31A}.

V2491~Cygni (Nova Cygni 2008 \#2) was discovered in outburst on 2008 April 10.8 UT with an unfiltered magnitude $m=7.7$ \citep{2008IAUC.8934....1N}.  Its nature was confirmed spectroscopically by \citet[H$\alpha$ FHWM $\sim4,500$~km~s$^{-1}$]{2008CBET.1334....1A} and was latter classified as a He/N nova \citep{2008CBET.1379....1H,2010arXiv1009.0822M}.  The rapid decline from maximum light classifies V2491~Cyg as a {\it very fast} nova.

The optical decline of V2491~Cyg exhibited a secondary maximum at day $\sim15$ \citep{2010arXiv1009.0822M}.  Such significant {\it re-brightenings} have been seen in a number of other novae \citep[V2362~Cyg and V1493~Aql;][]{2010AJ....140...34S}, but are unusual, and still poorly understood.  \citet{2009ApJ...694L.103H} have proposed magnetic activity as an additional energy source and a possible cause of the re-brightening, a requirement being a highly magnetised WD (a polar).

Following the outburst, it was discovered that there was a pre-existing X-ray source coincident with the position of V2491~Cyg \citep{2008ATel.1473....1I,2008ATel.1478....1I,2009A&A...497L...5I}.  This is only the second nova for which pre-outburst X-ray emission has been detected, after V2487~Oph \citep{2002Sci...298..393H}.  V2487~Oph was classified as a recurrent nova when a previous outburst was indentified from 1900 \citep{2009AJ....138.1230P}.

\citet{2008ATel.1514....1B} reported a variation in the $B$- and $V$-bands between 10 and 20 days after outburst, these had a period of 0.09580(5)~days and amplitude of $0.03-0.05$~mag.  This period has been taken by some authors to be the orbital period of the system.  However, Baklanov (private communication) have not made this claim and also note that this period became negligible as the system waned.  \citet{2010arXiv1004.0419R} reported similar $R$-band periodicity between day 28 and 34.

The first indication that V2491~Cyg may be recurrent in nature came from \citet{2008ATel.1485....1T} who noted the spectral similarities to the RNe U~Sco and V394~CrA early after outburst.  Additionally, \citet{2009ApJ...705.1056B} noted that the V2491~Cyg spectra were similar to the early spectra of RS~Oph and in particular those of M31N~2007-12b, a RS~Oph-class RN candidate in M31.

The progenitor system of V2491~Cyg was identified as USNO-B1.0~1223-042965 \citep{2008IBVS.5834....1H} and there is no evidence of previous outbursts at this position \citep{2008IBVS.5839....1J}.  \citet{2008IAUC.8938....2R} determined a reddening to V2491~Cyg of $E_{B-V}=0.43$ via OI line ratios.  \citet{2008CBET.1379....1H} used the CN MMRD relation \citep{1995ApJ...452..704D} with this reddening to determine $d=10.5$~kpc.  \citet{2010arXiv1009.0822M} independently derived both the extinction and distance to V2491~Cyg using the interstellar NaI line and the \citet{1987A&AS...70..125V} relation, finding $E_{B-V}=0.23\pm0.01$ and $d=14$~kpc.  \citet{Ribeiro2010} reported an ejecta morphology consisting of polar blobs and an equatorial ring, with expansion velocities of $\sim3,000$~km~s$^{-1}$ and an inclination of $80^{+3}_{-12}$~degrees, i.e. close to edge on.

\citet{2010MNRAS.401..121P} reported the results of extensive {\it Swift} X-ray and UV observations of the outburst, these data indicated that the WD in the system may be close to the Chandrasekhar mass.  By assuming the pre-outburst X-ray emission was due to accretion \citet{2010MNRAS.401..121P} deduced that V2491~Cyg would recur on longer than typical timescales, centuries rather than decades, if the WD mass was indeed high.  This paper noted in addition that X-ray flickering beginning at day 57 implied the resumption of accretion in the system.  \citet{2010MNRAS.401..121P} also reported that the $\sim0.1$~day period observed in the optical was not seen in the {\it Swift} UVOT or XRT data.  Interestingly, \citet{2010MNRAS.401..121P} show that the presence of a highly magnetic WD in the V2491~Cyg system is unlikely, in direct contradiction to the outburst mechanism proposed by \citet{2009ApJ...694L.103H}.

In this research note we address the nature of the progenitor system of V2491~Cyg and the $\sim0.1$~day periodicity observed early post-outburst.  Data collection and processing is described in Section~\ref{sec2} with analysis in Section~\ref{sec3}.  In Section~\ref{sec4} we present a discussion about the nature of the progenitor system.

\section{Observations}\label{sec2}

All additional data reported in this research note were obtained by the 2.0m robotic Liverpool Telescope \citep[LT;][]{2004SPIE.5489..679S} located at the Observatorio del Roque de Los Muchachos on the Canary Island of La Palma, Spain.  To investigate the source of the $\sim0.1$~day periodicity \citep{2008ATel.1514....1B,2010arXiv1004.0419R}, two series of 180 consecutive 60s $B$-band observations were taken by the LT on 2010 April 24 and 2010 September 11, $t=744$~days and $t=884$~days post-outburst respectively.  These series corresponded to 3.59 hours each, $\sim1.6$ times longer than the \citet{2008ATel.1514....1B} period.  We also make use of archival data from the Two Micron All Sky Survey \citep[2MASS;][]{2006AJ....131.1163S} in order to search for a potential infrared counterpart to the nova.  Pre-processing and photometry was performed on the LT and 2MASS data using various routines provided by IRAF.  The analysis of the near-IR data is described in Section~\ref{ir} and the optical in Section~\ref{period}.

\section {Results}\label{sec3}

\subsection {Infrared counterpart}\label{ir}

\begin{figure}
\center{\includegraphics[width=0.75\columnwidth]{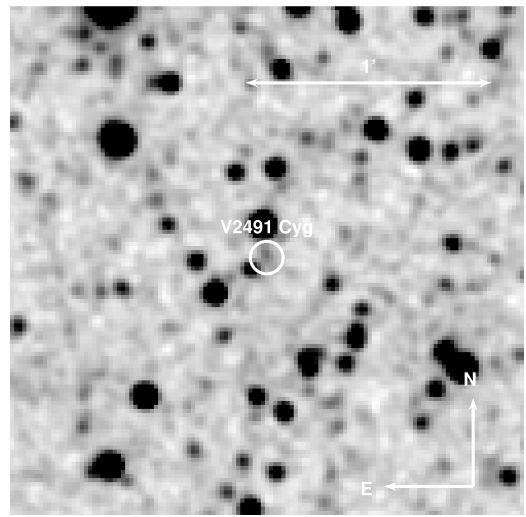}}
\caption{2MASS $J$-band image of the region around V2491~Cyg, the position of V2491~Cyg is marked by the circle.  There is a faint source at this position in the 2MASS $J$ data, also in the $H$ and $K_{S}$ data.\label{2MASS:J}}
\end{figure}

For a nova system containing a sub-giant or red giant secondary, one would expect the near-IR emission to be dominated by the secondary.  A solar-like sub-giant star ($M<1.4M_{\sun}$; the upper mass limit dictated by that of the WD primary for stable accretion) would be expected to have a luminosity $2\le M_{J}\le3$ \citep{2004ApJ...612..168P} with a red giant being even more luminous.  At distances of up to $\sim15$~kpc a red giant dominated system should be easily detectable within the 2MASS data; a sub-giant may be at or around the limit of detectability.  A search of the 2MASS catalogue at the position of V2491~Cyg reveals no objects within a reasonable radius.  However, upon inspection of the 2MASS images of the region there is a clear, but faint, source at the position of the nova, see Figure~\ref{2MASS:J}.  Photometry of this object, calibrated to the 2MASS catalogue for a large number of neighbouring stars, yields $m_{J}=16.7\pm0.2$,  $m_{H}=16.6\pm0.3$ and $m_{K_{S}}=16.7\pm0.6$.  The $J$-band photometry equates to an absolute magnitude of $M_{J}=1.2\pm0.2$ and $M_{J}=0.7\pm0.2$ for distances of $d=10.5$~kpc and $d=14$~kpc and reddenings $E_{B-V}=0.43$ and $E_{B-V}=0.23$ respectively.  For the $H$- and $K_{S}$-bands we find $M_{H}=1.2\pm0.3$ or $M_{H}=0.7\pm0.3$ and $M_{K_{S}}=1.4\pm0.6$ or $M_{K_{S}}=0.9\pm0.6$ for the two distances and extinctions, as above.  These are broadly consistent of the 2MASS luminosities of the RN U~Sco at quiescence \citep[$M_{J}=1.3\pm0.4$, $M_{H}=0.9\pm0.4$ and $M_{K_{S}}=0.9\pm0.4$;][]{1985MNRAS.213..443H,2010ApJS..187..275S}.

\subsection {0.1 day optical periodicity}\label{period}

\begin{figure}
\center{\includegraphics[width=0.75\columnwidth]{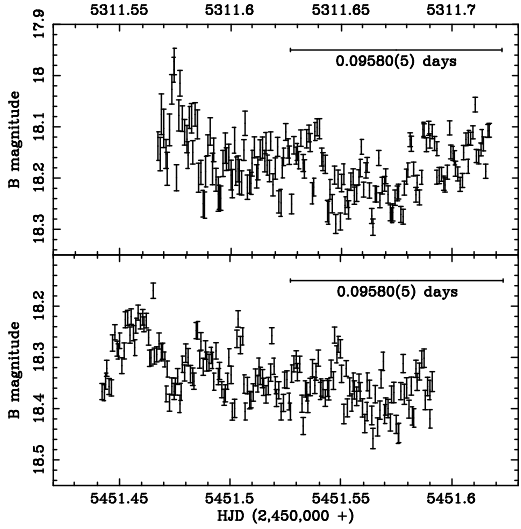}}
\caption{Liverpool Telescope $B$-band light curve of V2491~Cyg taken 744~days (top) and 884~days (bottom) post-outburst.  Each photometry run covers 3.59 hours, each observation lasted 60 seconds.  A ruler of length 0.09580(5)~days \citep[i.e. the period reported by][]{2008ATel.1514....1B} is shown on each plot.\label{LT_data}}
\end{figure}

The lightcurves from the LT $B$-band observing runs taken 744 and 884 days after the outburst of V2491~Cyg are shown in Figure~\ref{LT_data}, the mean magnitude during each run is shown in Table~\ref{photo_table}.  The photometric uncertainties quoted are dominated by the calibration of the data, and the brightness of the system decreased significantly between the two runs.  Such a decline may indicate that the object has not {\it quite} returned to quiescence, or it may be indicative of the system's variation at quiescence.  In addition to this decline, the $B$-band emission from the system is clearly variable, exhibiting an amplitude of $\Delta B\simeq0.2$~mag.  However, the variability is much more erratic than the almost {\it sine} curve-like variations seen by \citet{2008ATel.1514....1B}.  There is also no distinct period seen in these data; the first run has properties broadly consistent with a $\sim0.1$~day periodicity - possibly just coincidentally, but the second data set shows evidence of a much shorter $\sim0.025$~day period.  These data are too sparse to perform any meaningful period analysis.  However, it is clear that there is no strong eclipse signature seen in these data.  The eclipses predicted by \citet{Ribeiro2010} would be expected to be as deep as those seen in DQ~Her and U~Sco; $\sim1$ magnitude.  As such, eclipsing orbital periods $<0.15$~days may be ruled out.  These lightcurves are reminiscent of optical flickering seen once accretion had re-established following the 2006 outburst of RS~Oph \citep{2007MNRAS.379.1557W}.

\subsection {Optical photometry pre- and post-outburst}\label{optphot}

Whilst there is a vast array of optical data covering the outburst of V2491~Cyg, there is somewhat a lack of data for the system at, or close to, quiescence.  In Table~\ref{photo_table} we have consolidated both pre- and post-outburst data from a number of sources.  These sources include, data from the SDSS-II survey and the Asiago 1.82m using the AFOSC spectrograph/imager \citep[both as published in][]{2010arXiv1009.0822M}.  Comparison of the pre-outburst SDSS-II data with the later AFOSC and LT data indicates that V2491~Cyg is back (or very close) to quiescent levels by day $\sim800$ after outburst.  As such, we derive the outburst amplitude of $\Delta m_{V}\sim\Delta m_{R}\sim10$~magnitudes, which is similar to the outburst amplitude observed in U~Sco \citep{2010AJ....140..925S}.  The typical amplitudes observed in RS~Oph-class RNe is $\sim6-7$ magnitudes with those in CNe being in the range of $m_{V}\sim10$~magnitudes for the very slowest up to $m_{V}\sim17$~ magnitudes for the very fastest \citep{2008clno.conf....1W}.   These quiescent data can also, in principle, be used as an aid to determining the nature of the progenitor system, but one needs to be careful when dealing with the large uncertainties on the distance and extinction to V2491~Cyg.

\begin{table}
\begin{center}
\begin{tabular}{llll}
\hline\hline
Telescope & Filter & Time (days  & Magnitude \\
          &        & after peak) &\\
\hline
2MASS     & $J$    & -3,633 & $16.7\pm0.2$ \\
2MASS     & $H$    & -3,633 & $16.6\pm0.3$\\
2MASS     & $K_{S}$ & -3,633 & $16.7\pm0.6$\\
\hline
SDSS-II & $B$ & -6,895 & 18.3\\
SDSS-II & $R$ & -6,895 & 17.4\\
SDSS-II & $I$ & -6,895 & 16.9\\
\hline
AFOSC     & $V$      & +831 & 17.88\\
AFOSC     & $R$      & +831 & 17.49\\
AFOSC     & $I$      & +831 & 17.14\\
\hline
LT        & $B$      & +744 & $18.17\pm0.05$\\
LT        & $B$      & +884 & $18.34\pm0.05$\\
\hline
\end{tabular}
\end{center}
\caption{A summary of photometric observations of the V2491~Cyg progenitor system at or close to quiescence used in this work.  Data are taken from \protect{\citet{2010arXiv1009.0822M}} and this work.\label{photo_table}}
\end{table}

\begin{figure}
\begin{center}
\includegraphics[width=0.75\columnwidth]{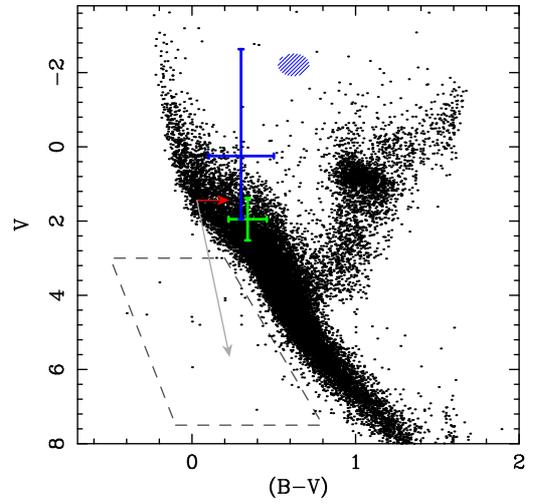}
\end{center}
\caption{Colour-magnitude diagram using {\it Hipparcos} data \protect{\citep{1997ESASP1200.....P}}.  The green point indicates the position of a quiescent U~Sco, the ellipse shows the location of a queiscent RS~Oph and the blue point shows the position of V2487~Oph.  The arrows respresent a quiescent V2491 Cyg: tail ($d=10.5$~kpc, $E_{B-V}=0.43$); red arrow-head ($d=14$~kpc, $E_{B-V}=0.23$); grey arrow-head ($d=2$~kpc, $E_{B-V}=0.23$; see Section~\ref{disc:dist}).  The grey dashed region indicates where most quiescent classical nova systems are found (Darnley et~al., in prep).\label{IvsV-I}}
\end{figure}

Shown in Figure~\ref{IvsV-I} is a $V$ {\it vs.} $B-V$ colour-magnitude diagrams populated with {\it Hipparcos} stars \protect{\citep{1997ESASP1200.....P}}.  The quiescent positions of the RN prototypes RS~Oph ({\it hatched} region) and U~Sco (green) and also the RN V2487 Oph (blue) are included for comparison.  The position of V2491~Cyg is shown by the base of the red arrow, assuming $d=10.5$~kpc and $E_{B-V}=0.43$; the red arrow head indicates the position of the system for $d=14$~kpc and $E_{B-V}=0.23$.  It is clear that the uncertainties on distance and extinction have little effect on the system's position in a colour-magnitude diagram.  This plot supports the results from the near-IR data that the V2491~Cyg system does not contain a red giant secondary.  However, the quiescent V2491~Cyg system lies remarkably close to the position of U~Sco.  Additionally, V2491~Cyg and V2487~Oph, both of which exhibited pre-outburst X-ray emission, lie within a similar region on this plot. 

\section{Discussion}\label{sec4}

\subsection{Nature of the system}

Many properties of V2491~Cyg are directly comparable with the RN U~Sco; the optical and near-IR luminosities at quiescence are almost identical, as is the outburst amplitude of the system.  As the spectral energy distributions of both V2491~Cyg and U~Sco are so similar, we must conclude that the composition of the system is similar; an accretion disk dominating the short wavelength emission the secondary being important at longer wavelengths.  Based with the quiescent photometric evidence we conclude that the V2491~Cyg system is likely itself to be a ``clone'' of the U~Sco system.  That is, highly inclinded (edge-on), containing a high mass WD primary, a sub-giant secondary and a bright accretion disk, and as such on this evidence alone we would expect the system to be recurrent on relatively short timescales.

Such a system is however incompatible with an orbital period as short as $0.1$~days as has been widely assumed for this system, as the radius of the sub-giant star would be larger than the orbital separation.  It is our opinion that any such periodicity seen so early after outburst is unlikely to be related - directly - to the orbital period of the system, as the system is expected to be entombed within the optically thick photosphere.  Indeed this period was observed before the emergence of the super soft source (SSS) which is widely attributed to the unveiling of the WD.  The SSS was only revealed at day $\sim40$ indicating that before this the central system was still likely to be obscured.  The nature of the optical spectra at this time \citep[P~Cygni profiles visible and an absence of high excitation lines;][]{2010arXiv1009.0822M} support this.  The U~Sco system is well known for its eclipsing nature.   However, during the 2010 outburst of U~Sco, the eclipses were masked until the ejecta had cleared the central system (around day~15; Schaefer et al. in prep).  Indeed these eclipses re-appeared three days {\it after} the SSS of U~Sco was revealed \citep[day 12;][]{2010ATel.2430....1S}.  As the ejecta velocities seen in U~Sco \citep[$\sim7600$~km~s$^{-1}$; see e.g.][]{2010ATel.2411....1A} are greater than those of V2491~Cyg, a simple approach would lead to the central system of the latter taking longer to be unveiled.  The early post-outburst light curve of U~Sco (days $0-9$) itself exhibited variation on a timescale much shorter than that system's orbital period.  As such, the early variability seen in both systems is most likely connected to the outburst itself.   The {\it quasi}-periodic variability seen in the LT light curve of V2491~Cyg at quiescence is most likely due to ``flickering'' of an accretion disk.

Given the similarities of this V2491~Cyg to U~Sco, we would expect the orbital period of V2491~Cyg to be similar to that of the former, i.e. days rather than hours.  Additionally, the larger distance derived by \citet{2010arXiv1009.0822M} has potential implications to the nature of the system.  \citet{2010MNRAS.401..121P} predicted a recurrence timescale for V2491~Cyg, based on a WD mass of $\sim1.3M_{\sun}$ and a distance of 10.5~kpc, of $>100$ years.  However, with V2491~Cyg as distant as 14~kpc, such a recurrence timescale decreases significantly to $<100$ years, on a par with the known recurrents, i.e. the inferred accretion rate is higher.

\subsection{Distance}\label{disc:dist}

A rather important caveat to the above argument is the distance to V2491~Cyg.  Both the \citet[10.5~kpc]{2008CBET.1379....1H} and \citet[14~kpc]{2010arXiv1009.0822M} distances are essentially derived via the same methodology; use of the MMRD relationship.  It is only in their estimation of the reddening \citep[by][for the former]{2008IAUC.8938....2R} that they differ.  It should be noted that both these distances are broadly consistent if one applies the $\sim0.6$ magnitude typical scatter around the MMRD.  It is also worth noting that the MMRD performs poorly for {\it very fast} novae like V2491~Cyg, the atypical light curve may also have affected the reliability.  \citet{2010ApJS..187..275S} has shown that the MMRD may perform poorly for recurrent novae significantly overestimating the distance in some cases (e.g. U~Sco).  V2491~Cyg is indeed not a typical nova, the re-brightening complicates the determination of the speed class, and may invalidate the MMRD for this nova.  In deriving the distance to V2491~Cyg of 14~kpc, \citet{2010arXiv1009.0822M} made a number of observations.  Given the Galactic coordinates of the system ($l=67.2$, $b=+4.4$~deg), the majority of the interstellar material giving rise to the observed reddening will lie within 2~kpc of the Sun.  At a distance of 14~kpc, V2491~Cyg is also unusually high above the Galactic plane (z=1.1~kpc) for a He/N nova.  However, both V477~Sct \citep{2006A&A...452..567M} and V2672~Oph \citep{2010MNRAS.tmp.1484M}, both He/N novae, also occupy such lofty positions above the plane.

We are given two alternatives.  Either, V2491~Cyg is a U~Sco-class RN and the MMRD works well for this particular recurrent; or, the system is much closer, $\sim2$~kpc.  However, there is one parameter of the nova outburst that is both distance and extinction independent, the outburst amplitude; for V2491~Cyg $\Delta m_{V}\sim10$~magnitudes.  Such a small amplitude is incompatible with a {\it very fast} CN and is only achievable for the slowest (faintest) of novae.  Using the outburst amplitude versus rate of decline relationship for CNe \citep[see][]{2008clno.conf....1W}, the amplitude for a nova with a decline time $t_{2}=4.8$~days \citep{2010arXiv1009.0822M} and inclination $\sim80$~degrees \citep{Ribeiro2010} is almost 16 magnitudes.  As such, the only scenarios compatible with such a small outburst amplitude are that V2491~Cyg is a (nearby) severely under-luminous and very unusual CN or that it is a (distant) U~Sco-class RN.  In either case V2491~Cyg is an interesting and important system worthy of further study.

\bibliographystyle{aa} 

\end{document}